\documentclass[prd,superscriptaddress,nofootinbib,amsmath,amssymb,aps,11pt]{revtex4-2}

\usepackage{bm}
\usepackage{amsfonts}
\usepackage{latexsym}
\usepackage[latin1]{inputenc}
\usepackage{graphicx}
\usepackage{amsmath}
\usepackage{palatino}
\usepackage{mathpazo}
\usepackage[normalem]{ulem}
\usepackage{xcolor}

\usepackage{booktabs}
\usepackage{dcolumn}

\usepackage{natbib}

\def\jnl@style{\it}
\def\aaref@jnl#1{{\jnl@style#1}}

\def\aaref@jnl#1{{\jnl@style#1}}

\def\aj{\aaref@jnl{AJ}}                   
\def\apj{\aaref@jnl{ApJ}}                 
\def\apjl{\aaref@jnl{ApJ}}                
\def\apjs{\aaref@jnl{ApJS}}               
\def\apss{\aaref@jnl{Ap\&SS}}             
\def\aap{\aaref@jnl{A\&A}}                
\def\aapr{\aaref@jnl{A\&A~Rev.}}          
\def\aaps{\aaref@jnl{A\&AS}}              
\def\mnras{\aaref@jnl{Mon.~Not.~Roy.~Astron.~Soc.}}             
\def\prd{\aaref@jnl{Phys.~Rev.~D}}        
\def\prc{\aaref@jnl{Phys.~Rev.~C}}  
\def\prl{\aaref@jnl{Phys.~Rev.~Lett.}}    
\def\qjras{\aaref@jnl{QJRAS}}             
\def\skytel{\aaref@jnl{S\&T}}             
\def\ssr{\aaref@jnl{Space~Sci.~Rev.}}     
\def\zap{\aaref@jnl{ZAp}}                 
\def\nat{\aaref@jnl{Nature}}              
\def\aplett{\aaref@jnl{Astrophys.~Lett.}} 
\def\apspr{\aaref@jnl{Astrophys.~Space~Phys.~Res.}} 
\def\physrep{\aaref@jnl{Phys.~Rep.}}      
\def\physscr{\aaref@jnl{Phys.~Scr}}       
\def\commat{\aaref@jnl{Comm.~Math.~Phys.}}              
\def\science{\aaref@jnl{Science}}               
\def\cqg{\aaref@jnl{Classical Quant.~Grav.}}            
\def\jpcs{\aaref@jnl{JPCS}}                                     
\def\ijmpd{\aaref@jnl{Int.~J.~Mod.~Phys.~D}}                    
\def\grg{\aaref@jnl{Gen.~Relat.~Gravit.}}               
\def\rpp{\aaref@jnl{Rep.~Prog.~Phys.}}          
\def\npa{\aaref@jnl{Nucl.~Phys.~A}}        
\def\lrr{\aaref@jnl{Living Rev.~Rel.}}                   
\def\jcap{\aaref@jnl{J.~Cosmology Astropart.~Phys.}}    
\def\rmp{\aaref@jnl{Rev.~Mod.~Phys.}}   

\begin{document}

\title{Rotating scalarized black holes: the role of the coupling}

\author{Kalin V. Staykov}
\affiliation{Department of Theoretical Physics, Faculty of Physics, Sofia University "St. Kliment Ohridski", Sofia 1164, Bulgaria}

\author{Daniela D. Doneva}
\affiliation{Theoretical Astrophysics, Eberhard Karls niversity of T\"ubingen, T\"ubingen 72076, Germany}

\author{Pedro G. S. Fernandes}
\affiliation{Institut f\"ur Theoretische Physik, Universit\"at Heidelberg, Philosophenweg 12 \& 16, 69120 Heidelberg, Germany}

\author{Stoytcho S. Yazadjiev}
\affiliation{Department of Theoretical Physics, Faculty of Physics, Sofia University "St. Kliment Ohridski", Sofia 1164, Bulgaria}
\affiliation{Institute of Mathematics and Informatics, 	Bulgarian Academy of Sciences, 	Acad. G. Bonchev St. 8, Sofia 1113, Bulgaria}

\begin{abstract}
   We perform an in-depth analysis of rotating scalarized black holes in scalar-Gauss-Bonnet gravity, where scalarization is induced by the spacetime curvature. Our results show that even for very large spins, the scalar charge can reach values comparable to those in the static limit, meaning it is not significantly suppressed. Consequently, curvature-induced scalarization can lead to non-GR signatures of similar magnitude in both static and rapidly rotating cases. For certain coupling parameters, these scalarized black hole solutions remain within the regime of validity of the effective field theory, where the theory has well-posed formulations.
\end{abstract}

\maketitle

\section{Introduction}
Recent advancements in gravitational wave detection \cite{LIGOScientific:2016lio,LIGOScientific:2018dkp,LIGOScientific:2021sio,Freire:2024adf} and very-long-baseline interferometry \cite{EventHorizonTelescope:2019ggy,EventHorizonTelescope:2020qrl,EventHorizonTelescope:2021dqv,EventHorizonTelescope:2022xqj}, have enabled tests of gravity with unprecedented accuracy, providing detailed insights into the strong-field regions around black holes and neutron stars. This progress underscores the need for precise theoretical predictions that can be directly compared with observations. As a result, the study of compact objects in theories beyond General Relativity (GR) and their astrophysical implications has attracted significant attention in recent decades.

A notable alternative to General Relativity is \emph{scalar-Gauss-Bonnet gravity} (sGB), a higher-curvature theory that, in the weak coupling regime, can also be interpreted as an effective field theory. Within this regime, the theory remains well-posed \cite{Kovacs:2020pns,Kovacs:2020ywu,Reall:2021voz}.
Motivations for sGB gravity are also deeply rooted in efforts to quantize gravity, as well as its unique status as one of the few higher-curvature modifications of GR that yields second-order field equations, thereby avoiding ghost instabilities. Unlike conventional scalar-tensor theories, sGB gravity evades no-scalar-hair theorems, allowing for the existence of black holes with nontrivial scalar hair \cite{Kanti:1995vq,Torii:1996yi,Pani:2009wy,Sotiriou:2013qea,Herdeiro:2015waa}. Additionally, the phenomenon of spontaneous scalarization, first identified in neutron stars \cite{Damour:1993hw}, also manifests in black holes within sGB theories, both induced by curvature \cite{Doneva:2017bvd,Silva:2017uqg,Antoniou:2017acq} and by spin \cite{Dima:2020yac,Berti:2020kgk,Herdeiro:2020wei,Fernandes:2024ztk}.
Despite this, rotating solutions \cite{Cunha:2019dwb,Collodel:2019kkx,Berti:2020kgk,Herdeiro:2020wei,Liu:2025mfn} or dynamical simulations \cite{East:2021bqk,AresteSalo:2022hua,Corman:2022xqg} in sGB theory are still scarce due to the higher complexity of the field equations in comparison with GR. Only a limited set of coupling functions between the scalar field and the Gauss-Bonnet invariant have been investigated in these settings, and for a relatively small range of parameters.

Previous studies \cite{Cunha:2019dwb} have demonstrated that certain coupling functions can strongly suppress the scalar field and significantly constrain the parameter space in which black holes with scalar hair exist. Consequently, most research on scalarization of rotating black holes has focused on the spin-induced case \cite{Dima:2020yac,Doneva:2020nbb,Berti:2020kgk,Herdeiro:2020wei,Elley:2022ept,Doneva:2023oww,Fernandes:2024ztk}. In this paper, motivated by recent findings in Ref. \cite{Doneva:2024ntw}, which suggest that modifying the original coupling employed in Ref. \cite{Cunha:2019dwb} can significantly expand the domain of existence of rotating scalarized solutions, we aim to investigate rapidly rotating scalarized black holes across a broad range of couplings between the scalar field and the Gauss-Bonnet invariant. Our goal is to determine whether stationary solutions align with the predictions of the $3+1$ simulations performed in Ref. \cite{Doneva:2024ntw} and, if so, to assess whether the scalar field remains highly suppressed at high spin. Furthermore, we explore whether these scalarized solutions are within the regime of validity of the effective field theory, where the theory is well-posed \cite{Kovacs:2020pns,Kovacs:2020ywu}.

The paper is organized as follows. In Sec. \ref{sec:sGB} we introduce sGB gravity, and discuss under which conditions scalarization is attainable. In Sec. \ref{sec:numerical} we describe the numerical method used to obtain rotating black hole solutions, which are analyzed in Sec. \ref{sec:results}. In Sec. \ref{sec:characteristics} we discuss different characteristics of the obtained scalarized solutions and explore whether they are within the regime of validity of the effective field theory. We conclude in Sec. \ref{sec:conclusion}.

\section{scalar-Gauss-Bonnet gravity and scalarization}
\label{sec:sGB}
The action for sGB theories is
\begin{eqnarray}\label{action}
S=\frac{1}{16\pi}\int d^4x \sqrt{-g} 
\Big[R - 2\nabla_\mu \varphi \nabla^\mu \varphi  + \alpha f(\varphi){\cal R}^2_{GB} \Big]  ,\label{eq:quadratic}
\end{eqnarray}
where $R$ denotes the Ricci scalar of the metric $g_{\mu\nu}$ and $\nabla_{\mu}$ is the covariant derivative. The coupling constant $\alpha$ has dimensions of $\mathrm{length}^2$ and ${\cal R}^2_{GB}$ denotes the Gauss-Bonnet invariant defined as ${\cal R}^2_{GB}=R^2 - 4 R_{\mu\nu} R^{\mu\nu} + R_{\mu\nu\alpha\beta}R^{\mu\nu\alpha\beta}$, where $R_{\mu\nu}$ and $R_{\mu\nu\alpha\beta}$ are the Ricci tensor and the Riemann tensor respectively.
The field equations derived from the action \eqref{eq:quadratic} are
\begin{equation}  \label{eq:feq_1}
  R_{\mu \nu} - \frac{1}{2}g_{\mu \nu}R = 2\left( \nabla_\mu \varphi \nabla_\nu \varphi - \frac{1}{2} g_{\mu \nu} \nabla_\sigma \varphi \nabla^\sigma \varphi + 2\alpha\, P_{\mu \alpha \nu \beta} \nabla^\alpha \nabla^\beta f(\varphi) \right),
  \end{equation}
where
\begin{equation*}
          P_{\alpha \beta \mu \nu}  = 2\, g_{\alpha [\mu}G_{\nu] \beta} + 2\, g_{\beta [\nu} R_{\mu] \alpha} -R_{\alpha \beta \mu \nu},
\end{equation*}
is the double-dual of the Riemann tensor (the square brackets denote anti-symmetrization), and the scalar field equation is
\begin{equation}
  \nabla_\mu \nabla^\mu \varphi = - \frac{\alpha}{4} \frac{df(\varphi)}{d\varphi}{\cal R}^2_{GB}.
  \label{eq:feq_2}
\end{equation}

The action \eqref{action} defines different classes of sGB theories controlled by the specific form of the coupling function $f(\varphi)$. In this paper, we are interested in the class of sGB theories exhibiting spontaneous curvature-induced scalarization \cite{Doneva:2017bvd,Silva:2017uqg,Antoniou:2017acq}. In this case, the coupling function obeys the conditions
\begin{equation}\label{eq:CouplingFuncCondition}
    \frac{df(0)}{d\varphi} = 0, \qquad \frac{d^2f(0)}{d\varphi^2} > 0.
\end{equation}
The first condition ensures that vacuum GR solutions, with a zero scalar field, remain valid solutions of the sGB field equations. The second condition renders these GR vacuum geometries susceptible to tachyonic instabilities once the spacetime curvature exceeds a certain threshold. In this scenario, thermodynamically favored and linearly stable spontaneously scalarized black holes can emerge. The simplest coupling function satisfying the above conditions and allowing spontaneous scalarization is $f(\varphi) = \varphi^2$. However, this choice leads to linearly unstable solutions \cite{Blazquez-Salcedo:2018jnn}. This instability can be mitigated through various approaches, including introducing nonlinearities in the coupling function \cite{Doneva:2017bvd,Blazquez-Salcedo:2018jnn,Silva:2018qhn,Minamitsuji:2018xde}, incorporating scalar field self-interactions \cite{Macedo:2019sem}, or coupling the scalar field to the Ricci curvature \cite{Antoniou:2021zoy}. In the present paper, we will adopt the first approach. Namely, we consider non-linearities in the coupling function, which we take to have the following form
\begin{equation} \label{eq:CouplingFunc}
    f(\varphi) = \frac{1}{2\beta} \left(1 - e^{-\beta \varphi^2}\right),
\end{equation}
where $\beta > 0$ is a dimensionless parameter. This parameter is useful for investigating the spectrum of solutions because, while $\alpha$ controls the onset of scalarization, $\beta$ is responsible for controlling deviations with respect to GR, as we will see below.

\section{Metric ansatz, numerical method, and physical quantities of interest}
\label{sec:numerical}
We are interested in studying stationary and axisymmetric configurations with two Killing vector fields, $k=\partial_t$ and $\Phi = \partial_\phi$. To this end, we employ the following circular metric ansatz in quasi-isotropic coordinates
\begin{equation}
  ds^2 = -f \mathcal{N}^2 dt^2 + \frac{g}{f} \left[ h \left(dr^2 + r^2 d\theta^2\right) + r^2 \sin^2\theta \left(d\phi - \frac{W}{r}\left(1-\mathcal{N}\right) dt\right)^2\right],
  \label{eq:metric}
\end{equation}
where $f$, $g$, $h$ and $W$ are dimensionless functions of $r$ and $\theta$, while $\mathcal{N}$ is related to the coordinate location of the event horizon, $r_H$, through 
\begin{equation*}
  \mathcal{N} \equiv \mathcal{N}(r) = 1-\frac{r_H}{r}.
\end{equation*}
Using the above metric anzatz, one can dimensionally reduce the field equations  \eqref{eq:feq_1}--\eqref{eq:feq_2} and derive a system of partial differential equations (PDEs) to be solved numerically. Since the equations are lengthy, we do not present here their explicit form and instead refer the reader to Ref. \cite{Fernandes:2022gde} for further details.

The boundary conditions are derived via a careful examination of the field equations at the boundaries of the domain of integration. For the sGB theories under consideration, the metric functions must obey the same boundary conditions as the Kerr solution does. Axisymmetry and regularity at the axis imply that
\begin{equation}
    \partial_\theta f = \partial_\theta g = \partial_\theta h = \partial_\theta W = \partial_\theta \varphi = 0
\end{equation}
for $\theta=0, \pi$.
Since the considered solutions are symmetric with respect to a reflection on the equatorial plane $\theta=\pi/2$, it is enough to consider only the range $\theta \in [0,\pi/2]$. By symmetry considerations, we obtain $\partial_\theta f = \partial_\theta g = \partial_\theta h = \partial_\theta W = \partial_\theta \varphi = 0$ for $\theta=\pi/2$.

In the numerical method we use the compactified radial coordinate
\begin{equation}
  x = 1-\frac{2r_H}{r},
  \label{eq:x}
\end{equation}
mapping the range $r\in [r_H, \infty)$ to
\begin{equation}
  x \in [-1,1].
\end{equation}
Regularity at the event horizon $r=r_H$, or equivalently, $x=-1$, leads to the following conditions\footnote{These conditions can be derived by solving the field equations order-by-order in an expansion around the event horizon.}
\begin{equation}
    \begin{aligned}
         f -2 \partial_x f = 0, \qquad g + 2 \partial_x g = 0, \qquad \partial_x h = 0, \qquad W - \partial_x W  = 0, \qquad \partial_x \varphi = 0.
    \end{aligned}
\end{equation}
Asymptotic flatness at infinity ($x=1$) imposes
\begin{equation}
    \begin{aligned}
        f = g = h = 1, \qquad \varphi = 0, \qquad \partial_x W + j \left( 1+\partial_x f \right)^2 = 0,
    \end{aligned}
\end{equation}
where $j$ is the dimensionless spin defined below in Eq. \eqref{eq:dimensionless-spin}.

To solve the PDEs, we used the code and method from Ref.~\cite{Fernandes:2022gde}, to which the reader is referred for details. Namely, a pseudospectral method is implemented combined with the Newton-Raphson root-finding algorithm \cite{numericalMethodsDias}. The code has been previously used to successfully generate different types of numerical solutions, e.g., in Refs. \cite{Eichhorn:2023iab,Eichhorn:2025aja,Fernandes:2024ztk,Burrage:2023zvk,Fernandes:2022kvg,Lai:2023gwe,Guo:2023mda,Xiong:2023bpl,Ye:2024pyy,Guo:2024bkw,Guo:2024lck}. The metric functions and the scalar field are each expanded in a spectral series, with resolutions $N_x$ and $N_\theta$ in the radial ($x$) and angular ($\theta$) coordinates, respectively. The spectral series used for each of the functions
$\mathcal{F}^{(k)}=\{f,g,h,W,\varphi\}$ is
\begin{equation}
  \mathcal{F}^{(k)} = \sum_{n=0}^{N_x-1} \sum_{m=0}^{N_\theta-1} c_{nm}^{(k)} T_n(x) \cos \left(2m\theta\right),
\label{eq:spectralexpansion1}
\end{equation}
where $T_n(x)$ denotes the $n^{\rm th}$ Chebyshev polynomial, and $c_{nm}^{(k)}$ are the spectral coefficients to be determined by the root-finding method. The angular boundary conditions are automatically satisfied with this spectral expansion, and therefore are not explicitly imposed in the numerical method. After extensive testing, we fixed the resolution at $N_x=60$, and $N_\theta=12$, and convergence is declared once the norm of the change in the spectral coefficients between successive iterations is less than a specified tolerance taken to be of order $\mathcal{O}\left( 10^{-10} \right)$. For very large values of spin, we observed a decrease of accuracy as expected from the results of Ref. \cite{Fernandes:2022gde}. This is because we approach extremality/criticality and the metric ansatz \eqref{eq:metric} together with the aforementioned boundary conditions should be adapted in that limit.

The Komar angular momentum $J$ and mass $M$ can be extracted from the asymptotic fall-offs of the metric functions
\begin{equation}
    g_{tt} \sim -1 + \frac{2M}{r} + \mathcal{O}\left(r^{-2}\right), \qquad g_{\phi t} \sim \frac{2J}{r^2}\sin^2 \theta + \mathcal{O}\left(r^{-3}\right).
\end{equation}
The scalar field decays as
\begin{equation}
    \varphi \sim \frac{D}{r} + \mathcal{O}\left(r^{-2}\right),
\end{equation}
where $D$ is the scalar charge of the solution. 
We define the dimensionless spin $j$ as
\begin{equation}
    j \equiv J/M^2.
    \label{eq:dimensionless-spin}
\end{equation}
The Hawking temperature, $T_H$, event horizon area $A_H$, and angular velocity of the horizon $\Omega_H$ are given by
\begin{equation} \label{eq:phys_quantities}
    T_H = \frac{1}{2\pi r_H} \frac{f}{\sqrt{g h}}, \qquad A_H = 2\pi r_H^2 \int_0^\pi \frac{g \sqrt{h}}{f} \sin \theta \mathrm{d}\theta , \qquad \Omega_H = \frac{W}{r_H}, \qquad \mathrm{at} \qquad r=r_H.
\end{equation}
The entropy of sGB black holes is different from the Bekenstein-Hawking entropy, and is given by \cite{Fernandes:2022gde}
\begin{equation}\label{eq:entropy}
    S_H = \frac{A_H}{4} + \frac{\alpha}{2}\int_H d^4x\sqrt{ h}f(\varphi){\cal R}^2_{GB},
\end{equation}
where the integral is performed on the horizon.

Black hole solutions of the theory \eqref{action} obey a Smarr-type relation, which takes the following form \cite{Fernandes:2022gde}
\begin{equation}
    M = 2T_H S_H + 2\Omega_H J + M_s,
\end{equation}
where $T_H$ is the temperature on the horizon and $M_s = \frac{1}{2\pi}\int d^3x\sqrt{-g} \frac{f(\varphi)}{f'(\varphi)}\Box \varphi$. The fulfillment of this relation is an additional criterion we used for monitoring the accuracy of the code. For the most extreme solutions this equality is fulfilled with an accuracy of $10^{-4}$, but it is generally a few orders of magnitude lower for most of the solutions.

\section{Results}
\label{sec:results}
We have generated solutions for different values of the parameter $\beta$ in the coupling function \eqref{eq:CouplingFunc} and analyzed their properties. While Ref. \cite{Cunha:2019dwb} shows that $\beta = 6$ severely restricts the domain of existence for rotating scalarized black holes, we show that such low values of $\beta$ mostly produce solutions that lie beyond the weak coupling regime. 
Increasing $\beta$ suppresses the scalar field and makes the black holes weakly coupled. In addition, high $\beta$ significantly expands the range of parameters where scalarized solutions exist, allowing for configurations with high spin, in agreement with the results of Ref. \cite{Doneva:2024ntw}.

\subsection{Domain of existence and scalar charge}
For the employed coupling function \eqref{eq:CouplingFunc}, the Kerr metric, together with $\varphi = 0$, remains a valid solution of the field equations \eqref{eq:feq_1},\eqref{eq:feq_2} and exists for spin values up to $j \leq 1$. However, black holes can undergo scalarization within a restricted region of the parameter space, in the regime where curvature is high, which occurs for smaller black holes (compared to the coupling $\alpha$). The domain of existence is bounded on one side by the bifurcation line, where the Kerr solution becomes unstable under linear perturbations and scalarized black holes emerge. On the other side, it is limited by the critical line, beyond which solutions cease to exist due to violation of the regularity conditions for the metric functions and the scalar field at the horizon. Static and spherically symmetric black holes with a nontrivial scalar field exist for $M/\sqrt{\alpha} \lesssim 0.587$ \cite{Doneva:2017bvd,Cunha:2019dwb}. This threshold shifts to higher $M/\sqrt{\alpha}$ with the increase of the dimensionless angular momentum $j$.  For the chosen coupling, the bifurcation line is independent of $\beta$. The critical line is, however, highly dependent on $\beta$, and we will show that it shifts towards lower masses for larger $\beta$. Thus, the domain of existence of scalarized black holes expands with an increase of $\beta$.

\begin{figure}[htb]
	\includegraphics[width=0.95\textwidth]{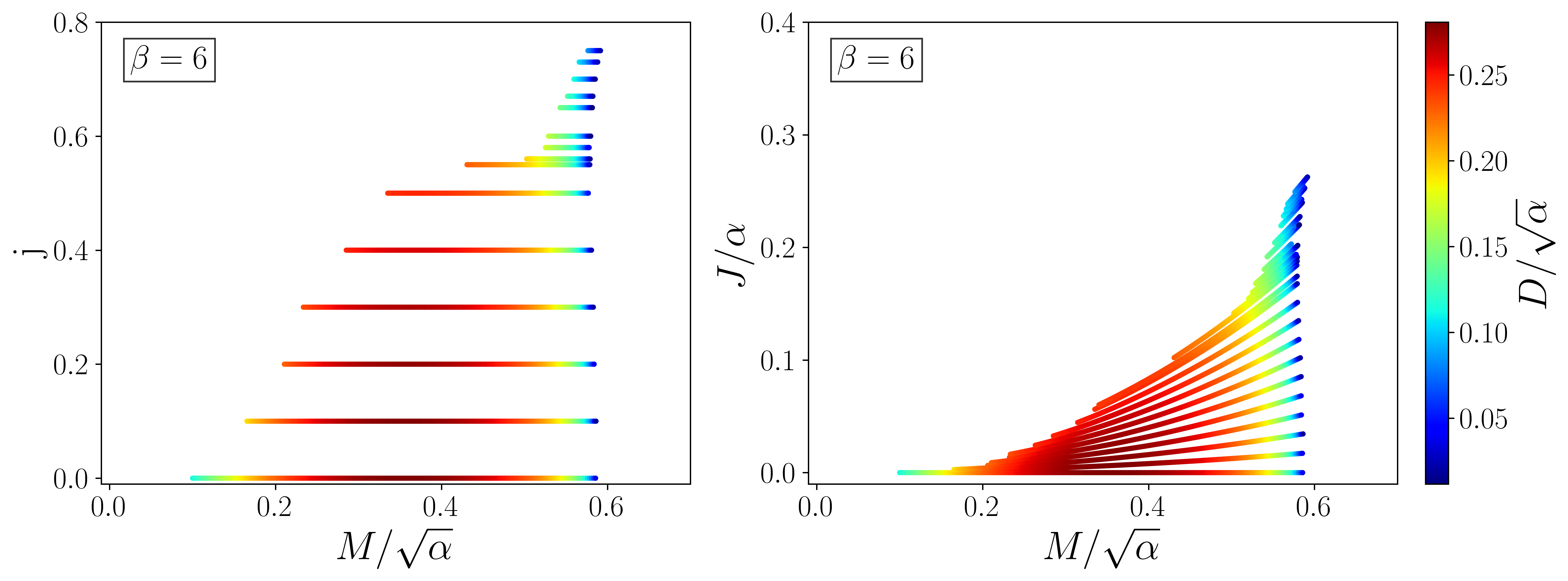}
	\caption{Case $\beta = 6$. \textit{Left panel}: The domain of existence of black hole solutions in the $\left(M/\sqrt{\alpha},j\right)$ plane. \textit{Right panel}: The domain of existence of black hole solutions in the $\left(M/\sqrt{\alpha},J/\alpha\right)$ plane. The scalar charge $D/\sqrt{\alpha}$ is given in color gradient on both panels.} \label{fig:b6}
\end{figure}

In Fig. \ref{fig:b6}, we present the case of $\beta = 6$ which has previously been studied in the literature \cite{Cunha:2019dwb}. The two panels depict the domain of existence for scalarized black holes in the $\left(M/\sqrt{\alpha},j\right)$ and $\left(M/\sqrt{\alpha},J/\alpha\right)$ planes.  Solutions are computed along sequences of constant $j$ and the colour gradient represents the strength of the scalar charge $D/\sqrt{\alpha}$. As can be seen in the left panel, the maximum $j$ we could obtain is $j = 0.76$. The left boundary of the domain of existence (the critical line), moves rapidly to higher masses with the increase of $j$. Therefore, the constant-$j$ solution sequences span a decreasing range of $M/\sqrt{\alpha}$, especially above $j\sim 0.6$. 
The domain of existence in the $\left(M/\sqrt{\alpha},J/\alpha\right)$ plane (the right panel in Fig. \ref{fig:b6}) is in a very good agreement with Fig. 1 in Ref. \cite{Cunha:2019dwb}. The only major difference is that while in Ref. \cite{Cunha:2019dwb} the maximum angular momentum reaches $J/\alpha = 0.4$, our models are restricted to $J/\alpha \sim 0.3$. This is a due to the numerical difficulties in obtaining black hole solutions for very high $j$, especially when they span a rapidly decreasing range of $M/\sqrt{\alpha}$.

For static or slowly rotating black holes, the scalar charge along constant $j$ sequences starts from zero at the bifurcation point (the right boundary of the domain), increases with the decrease of $M/\sqrt{\alpha}$ until a maximum is reached, and then starts to decrease again in the $M/\sqrt{\alpha} \rightarrow 0$ limit.  The branches of solutions get shorter for higher $j$, though, and they start being terminated before a maximum of $D/\sqrt{\alpha}$ is reached. In addition, the overall strength of the scalar field is suppressed as rotation increases. As a result, the maximum value of $D/\sqrt{\alpha}$ for rapidly rotating black holes differs by an order of magnitude from the typical $D/\sqrt{\alpha}$ values in the static case.

\begin{figure}[htb]
	\includegraphics[width=0.95\textwidth]{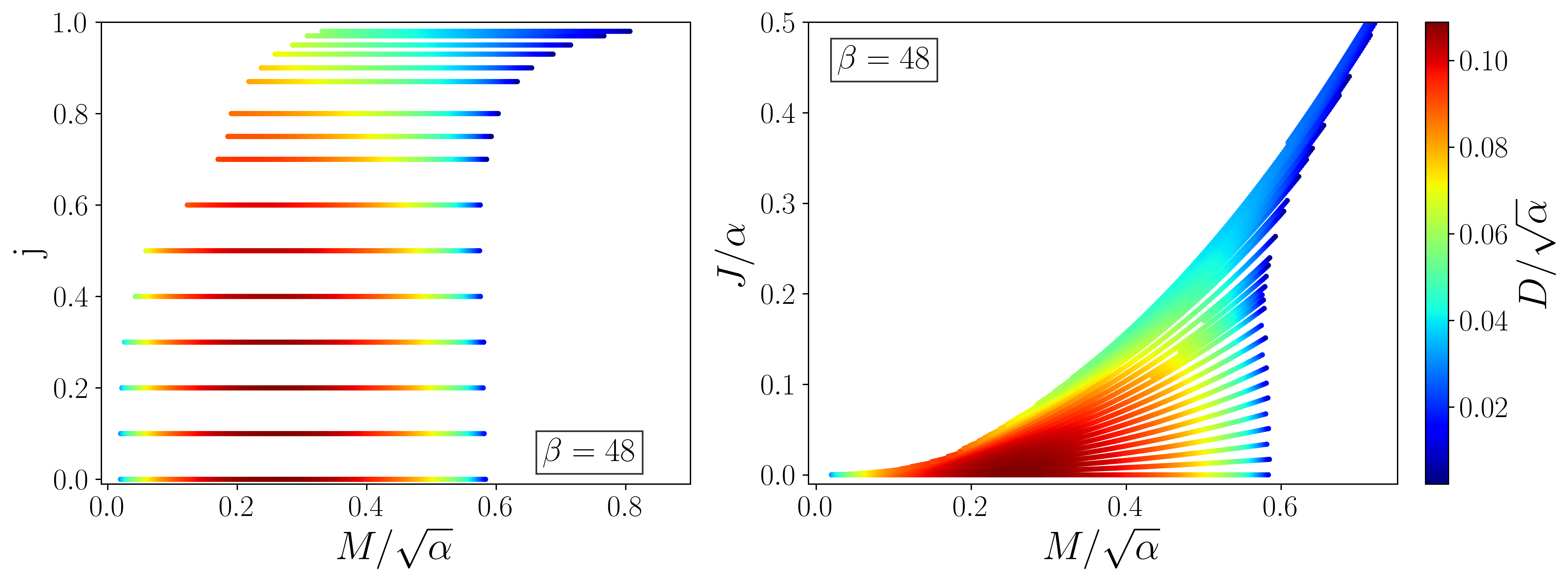}
	\caption{Case $\beta = 48$. \textit{Left panel}: The domain of existence of black hole solutions in the $\left(M/\sqrt{\alpha},j\right)$ space. \textit{Right panel}: The domain of existence of black hole solutions in the $\left(M/\sqrt{\alpha},J/\alpha\right)$ space. The scalar charge $D/\sqrt{\alpha}$ is given in color gradient on both panels }	\label{fig:b48}
\end{figure}

\begin{figure}[htb]
    \includegraphics[width=0.95\textwidth]{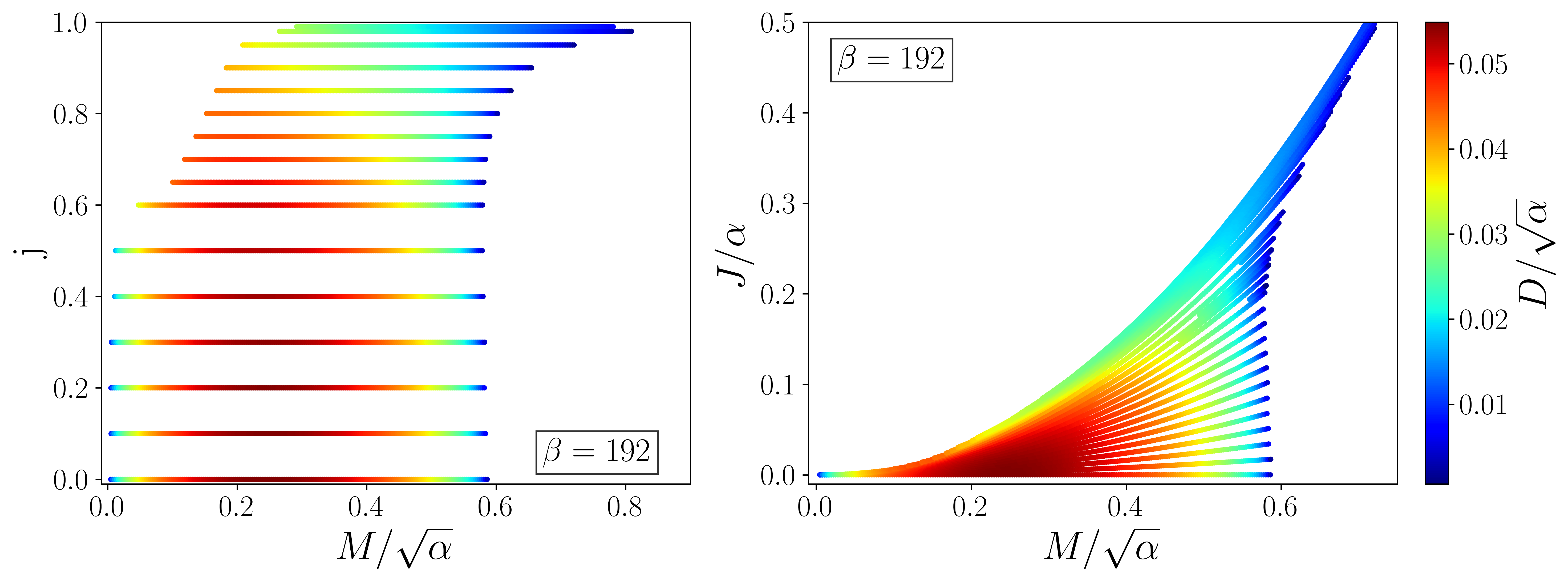}
	\caption{Case $\beta = 192$.  \textit{Left panel}: The domain of existence of black hole solutions in the $\left(M/\sqrt{\alpha},j\right)$ space. \textit{Right panel}: The domain of existence of black hole solutions in the $\left(M/\sqrt{\alpha},J/\alpha\right)$ space. The scalar charge $D/\sqrt{\alpha}$ is given in color gradient on both panels.}	\label{fig:b192}
\end{figure}

In Figs. \ref{fig:b48} and \ref{fig:b192}, we present our results for $\beta = 48$ and $\beta = 192$, respectively. In both cases, we were able to obtain black holes with significantly higher angular momenta, approaching $j \rightarrow 1$. In addition, the constant $j$ sequences span a much wider range of $M/\sqrt{\alpha}$ especially for rapid rotation. Nevertheless, for all considered values of $\beta$, we observed shortening of the sequences with respect to the static solutions, above roughly $j \sim 0.6$. An interesting question is whether the Kerr limit $j = 1$ could be surpassed for a given value of $\beta$. Despite an extensive numerical search, we were unable to obtain black holes with $j>1$.

The scalar charge $D/\sqrt{\alpha}$, colour-coded in Figs. \ref{fig:b48} and \ref{fig:b192}, decreases milder with rotation as $\beta$ increases. For example, for $\beta = 192$, the low mass rapidly rotating black hole solutions with $j \rightarrow 1$ have scalar charges roughly twice smaller compared to the maximum $D/\sqrt{\alpha}$ in the static limit. The main reason is that, for each $j$, it is possible to reach much lower values of $M/\sqrt{\alpha}$ compared to the $\beta=6$ case. Therefore, the spacetime curvature close to the horizon and thus the scalar field strength, increases. This trend is expected to continue with a further increase of $\beta$. Interestingly, despite the relatively large span of $M/\sqrt{\alpha}$ observed for rapidly rotating black holes with higher $\beta$, a maximum of $D/\sqrt{\alpha}$ is not reached contrary to the static and low $j$ sequences.

One can argue that despite the fact that large $\beta$ models exhibit only a mild decrease of $D/\sqrt{\alpha}$ as rotating increases, the absolute value of the scalar charge drops consistently for all solutions (both rotating and nonrotating) as $\beta$ is increased. This potentially limits detectability. As we will show below, though, a coupling parameter as low as $\beta=6$ leads to solutions which are not within the weak coupling regime. So if we want to treat sGB gravity as an effective field theory, one should consider much higher $\beta$, such as the ones considered in Figs. \ref{fig:b48} and  \ref{fig:b192}.

\section{Physical characteristics of the scalarized black holes and weak coupling condition}
\label{sec:characteristics}
\begin{figure}[htb]
    \includegraphics[width=0.95\textwidth]{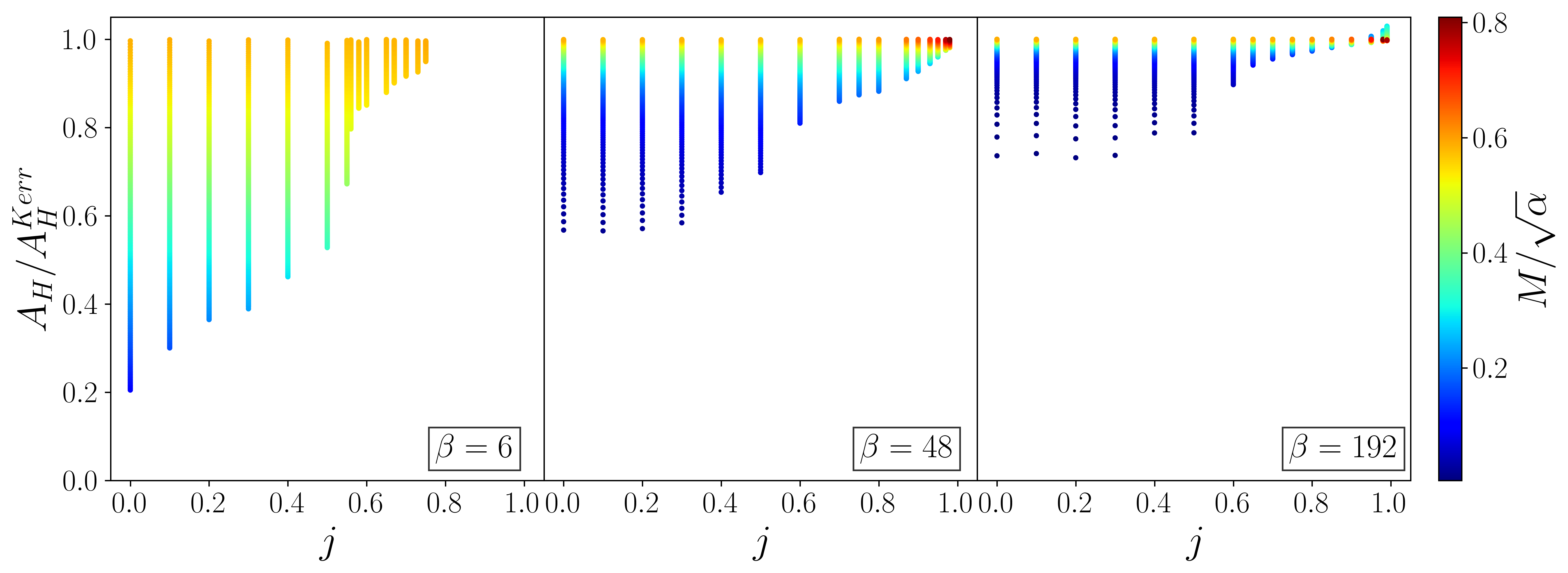}
	\caption{ The reduced area of the horizon $A_H/A_H^{Kerr}$, where $A_H^{Kerr} = 8\pi(M^2 + \sqrt{M^4 + J^2})$,  for different values of the dimensionless spin $j$. From left to right $\beta = 6$, $\beta = 48$, and $\beta = 192$. The colour gradient depicts the normalized black hole mass $M/\sqrt{\alpha}$.} 	\label{fig:Ah}
\end{figure}

In Fig. \ref{fig:Ah} we present the value of the horizon area $A_H$ normalized to the value of the horizon area of a Kerr black hole with the same mass and angular momentum, $A_H/A_H^{Kerr}$, for different dimensionless spins $j$. Here $A_H$ is calculated according to Eq. \eqref{eq:phys_quantities} and $A_H^{Kerr} = 8\pi(M^2 + \sqrt{M^4 + J^2})$. The colour gradient depicts the normalized black hole mass $M/\sqrt{\alpha}$. From left to right we have $\beta = 6$, $\beta = 48$, and $\beta = 192$. The observed behavior is consistent with the one reported in Ref. \cite{Cunha:2019dwb}. For a fixed $j$, when we move from the bifurcation line to the critical line (top to bottom in the figure), the area of sGB black holes decreases compared to that of Kerr black holes, except for $\beta=192$ and spins close to $j=1$, where the Kerr black hole has a smaller horizon area (the top-right corner of the rightmost panel). When comparing the cases for different $\beta$, it becomes evident that the difference in horizon area between scalarized and Kerr black holes decreases as $\beta$ increases. This is expected, as a higher $\beta$ brings the theory closer to GR. 

\begin{figure}[htb]
    \includegraphics[width=0.95\textwidth]{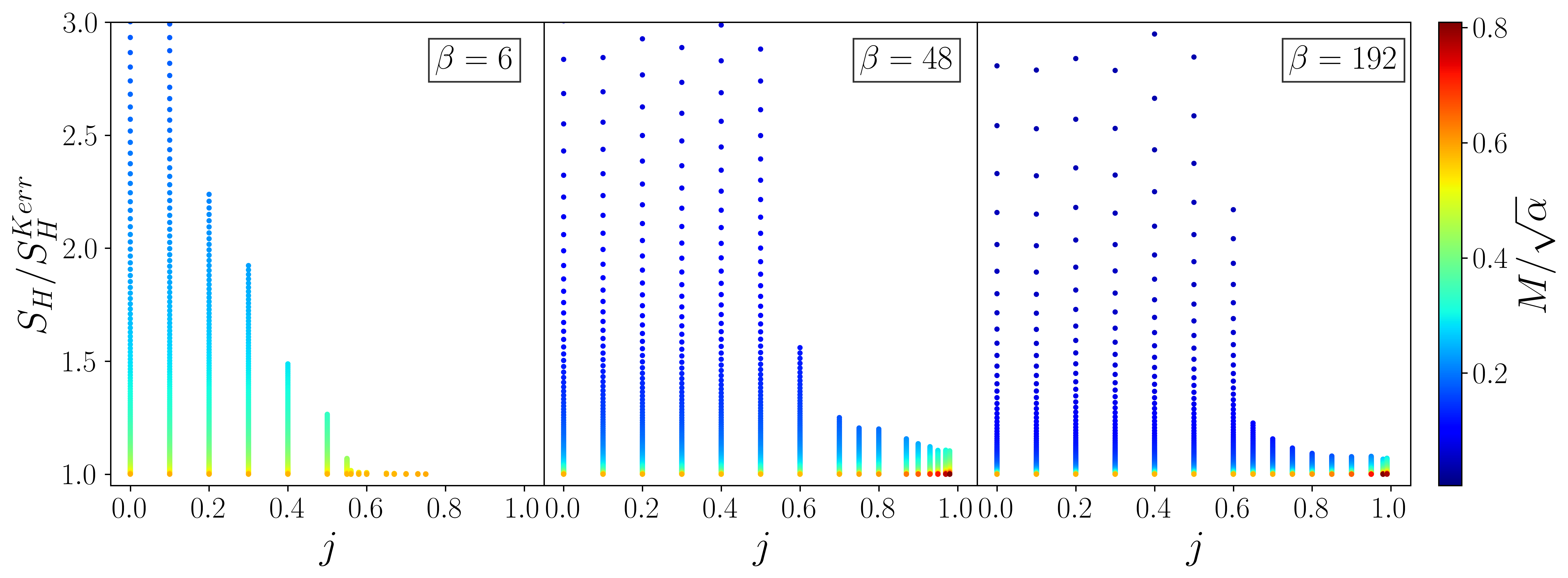}
	\caption{ The reduced black hole entropy $S_H/S_H^{Kerr}$, where $S_H^{Kerr} = A_H^{Kerr}/4$, for different values of the dimensionless spin $j$. From left to right $\beta = 6$, $\beta = 48$, and $\beta = 192$. The colour gradient depicts the normalized black hole mass $M/\sqrt{\alpha}$.} \label{fig:Sh}
\end{figure}

The reduced entropy $S_H/S_H^{Kerr}$ for different dimensionless spins $j$ is presented in Fig. \ref{fig:Sh}. Here $S_H$ is calculated according to Eq. \eqref{eq:entropy} and $S_H^{Kerr} = A_H^{Kerr}/4$.
The ratio $S_H/S_H^{Kerr}$ increases monotonically as we depart from the bifurcation line and it is always greater than unity despite the fact that the horizon area of the scalarized black holes is generally smaller than Kerr. This is due to the additional scalar field contribution in the expression for the entropy in sGB gravity, see Eq. \eqref{eq:entropy}. Therefore, scalarized black holes are entropically preferred for all values of $\beta$ and the whole domain of existence.

\begin{figure}[htb] 
	\includegraphics[width=0.95\textwidth]{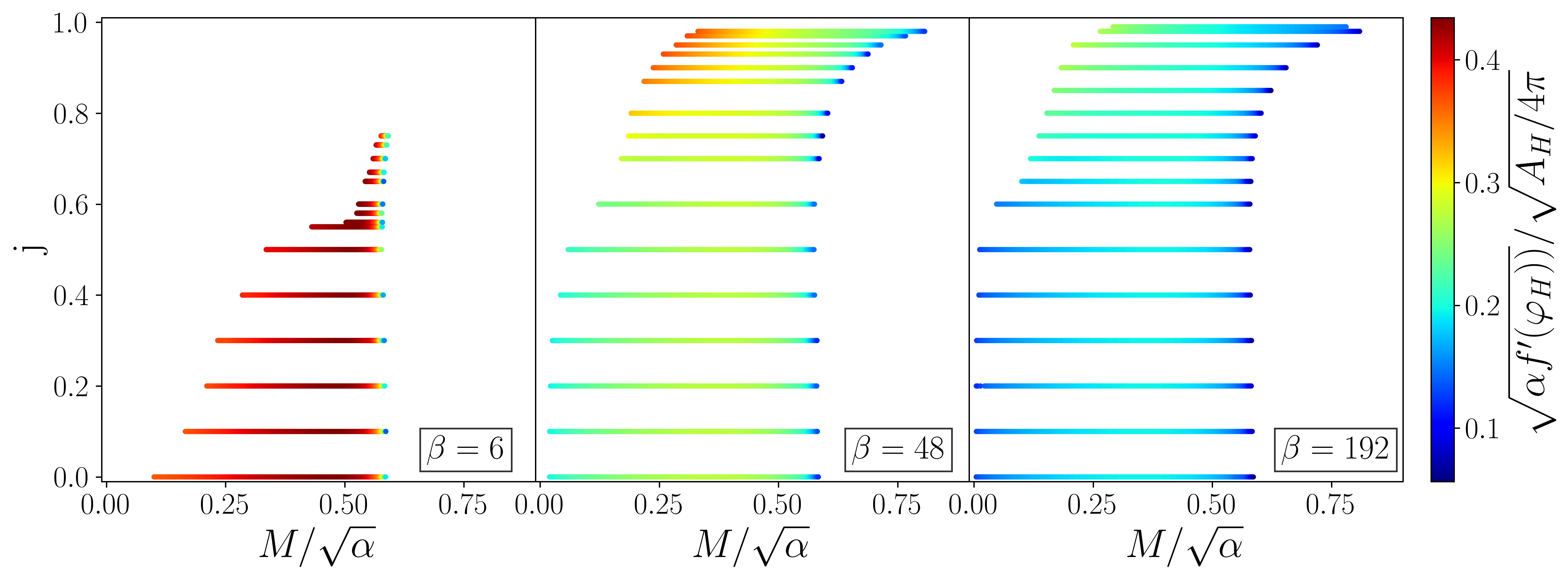}
	\caption{The domain of existence of scalarized black hole solutions in the $\left(M/\sqrt{\alpha},j\right)$ plane for different values of $\beta$. The weak coupling condition, defined by \eqref{eq:WCC}, is given in colour gradient in each panel.}	\label{fig:WCC}
\end{figure}

Finally, we comment on an important characteristic of the solutions, namely the weak coupling condition. Since only in this regime sGB gravity can be considered as a valid effective field theory, it is important to check whether the obtained solutions fall into this category. Moreover, a well-posed formulation of sGB gravity has, up to now, only been achieved in this regime \cite{Kovacs:2020pns,Kovacs:2020ywu,Reall:2021voz}. The weak coupling condition, and its violation for dynamical simulations  within sGB gravity, was discussed in detail in \cite{Ripley:2019hxt,AresteSalo:2022hua,AresteSalo:2023mmd,Doneva:2023oww,R:2022hlf,Thaalba:2023fmq,Thaalba:2024crk}. In this work, to assess the weak coupling condition, we employ a rough estimate based on the requirement that the characteristic length scale of the system, associated with the spacetime curvature, should be much larger than the length scale related to the coupling between the scalar field and the Gauss-Bonnet invariant. Namely, we require that
\begin{equation} \label{eq:WCC}
    \sqrt{A_H/4\pi} \gg \sqrt{\alpha f'(\varphi_H)} 
\end{equation}
where $\varphi_H$ is the value of the scalar field on the equatorial plane $\theta=\pi/2$, and $\sqrt{A_H/4\pi}$ serves as a measure of the gravitational radius of the black hole.

A contour plot of the weak coupling condition for the several couplings is shown in Fig. \ref{fig:WCC}. As illustrated, solutions for $\beta = 6$ are  strongly coupled both in the static and rapidly rotating regimes, with the exception of a small region close to the bifurcation line where the scalar field is very weak. Increasing $\beta$ generally reduces the scalar field and thus mitigates the violations of the weak coupling. In addition, a general observation is that rapid rotation tends to exacerbate the violation of the weak coupling condition. Nevertheless,  for $\beta = 192$, we observe that solutions across the entire parameter range lie on the border of being weakly coupled. Therefore, considering larger values of $\beta$ is not merely a mathematical strategy to generate scalarized solutions over a broader range of parameters. Rather, it is a physical necessity to bring the black holes into the weak coupling regime, where sGB gravity serves as a valid effective field theory and time evolution can be properly studied.

\section{Conclusion}
\label{sec:conclusion}
In this paper, we conducted a systematic study of rotating scalarized black holes in sGB gravity, considering various couplings between the scalar field and the Gauss-Bonnet invariant. For these solutions, the scalarization mechanism is driven by spacetime curvature. Previous studies \cite{Cunha:2019dwb} showed a strong suppression of the domain of existence and a reduction in the scalar field strength as rotation increases. As a result, spontaneously scalarized rotating black holes were primarily considered astrophysically relevant in the context of spin-induced scalarization \cite{Dima:2020yac,Doneva:2020nbb,Berti:2020kgk,Herdeiro:2020wei,Fernandes:2024ztk}. However, the results in Ref. \cite{Doneva:2024ntw} suggested that modifying the original coupling in Ref. \cite{Cunha:2019dwb} could significantly alter the domain of existence. Our aim was to explore this issue further and demonstrate that curvature-induced scalarized black holes can exist and present non-negligible deviations from the Kerr metric even for rapid rotation.

We have demonstrated that this is indeed the case. Specifically, for high values of $\beta$ in the coupling function \eqref{eq:CouplingFunc}, we found that the domain of existence expands, and the scalar charge is no longer highly suppressed, even for rapid rotation. In fact, the scalar charge becomes comparable to that of static black holes, opening up the possibility for interesting astrophysical implications. Additionally, we systematically explored the area of the event horizon and the entropy. Scalarized black holes generally have smaller areas than their Kerr counterparts, but they always exhibit higher entropy, making them thermodynamically favorable.

It is important to emphasize that varying the parameter $\beta$ is not simply an exploration of the parameter space. sGB gravity is typically motivated as an effective field theory, and in order for it to serve as a valid truncation of a more fundamental theory, it must remain within the weak coupling regime. Furthermore, this is the regime in which a well-posed formulation of sGB has been established. We demonstrated that for low values of $\beta$ (such as $\beta = 6$ employed in \cite{Cunha:2019dwb}), the scalar field becomes strongly enhanced, causing the black hole solutions to significantly violate the weak coupling condition. For higher values of $\beta$ (such as $\beta = 192$), the solutions lie on the edge of the weak coupling regime, even for the most extreme rotations we considered. Therefore, the overall conclusion is that curvature-induced scalarized black holes can exist in both static and rapidly rotating regimes, with deviations from GR that are non-negligible, while the weak coupling condition is satisfied.

As a direction for future research, it would be interesting to investigate whether the main conclusions of this paper hold for the theory presented in Ref.~\cite{Eichhorn:2023iab}, where curvature-induced scalarization occurs at supermassive scales, as well as for the theory recently proposed in Ref. \cite{Eichhorn:2025aja}, where curvature-induced scalarization is achieved without the explicit introduction of a scalar field in the action.

\section{Acknowledgements}
K.S. and S.Y.  are supported  by the European Union-NextGenerationEU, through the National Recovery and Resilience Plan of the Republic of Bulgaria, project No. BG-RRP-2.004-0008-C01. DD  acknowledges financial support via an Emmy Noether Research Group funded by the German Research Foundation (DFG) under grant no. DO 1771/1-1. We acknowledge Discoverer PetaSC and EuroHPC JU for awarding this project access to Discoverer supercomputer resources. P.~F.~acknowledges support by a grant from Villum Fonden under Grant No.~29405. This work is funded by the Deutsche Forschungsgemeinschaft (DFG, German Research Foundation) under Germany's Excellence Strategy EXC 2181/1 - 390900948 (the Heidelberg STRUCTURES Excellence Cluster).

\bibliography{references}

\end{document}